\documentclass[]{spie}  %>>> use for US letter paper
\usepackage[dvips]{graphicx}

\title{Deconvolution with a spatially-variant PSF} 

\author{Tod R. Lauer\supit{a}
\skiplinehalf
\supit{a}National Optical Astronomy Observatory, P.O. Box 26732,
Tucson, AZ 85726}

\authorinfo{Further author information: E-mail: lauer@noao.edu, Telephone: (520) 318-8290}
\begin{document} 
  \maketitle 

%%%%%%%%%%%%%%%%%%%%%%%%%%%%%%%%%%%%%%%%%%%%%%%%%%%%%%%%%%%%% 
\begin{abstract}
Application of deconvolution algorithms to astronomical images
is often limited by variations in PSF structure over the domain
of the images.  One major difficulty is that Fourier methods
can no longer be used for fast convolutions over the entire images.
However, if the PSF is modeled as a sum of orthogonal functions that are
individually constant in form over the images, but whose relative amplitudes
encode the PSF spatial variability, then separation of variables
again allows global image operations to be used.  This approach is
readily adapted to the Lucy-Richardson deconvolution algorithm.
Use of the Karhunen-Lo\`eve transform allows for a particularly
compact orthogonal expansion of the PSF.  These techniques
are demonstrated on the deconvolution of Gemini/Hokupa'a adaptive
optics images of the galactic center.
\end{abstract}

%>>>> Include a list of keywords after the abstract 


%%%%%%%%%%%%%%%%%%%%%%%%%%%%%%%%%%%%%%%%%%%%%%%%%%%%%%%%%%%%%
\section{INTRODUCTION}
\label{sect:intro}  % \label{} allows reference to this section

The deconvolution of astronomical images becomes tricky when the structure
of the point-spread function (PSF) varies significantly over the domain
of interest.  A constant PSF is generally not a formal requirement for
most deconvolution algorithms, but it is a profoundly useful
simplification.  With a constant PSF, one can quickly perform convolutions
over the entire image in the Fourier domain, and the difficult problem
of trying to estimate the PSF structure as it varies from point to
point is completely avoided.  Unfortunately, real astronomical images
often do have spatially-variant PSFs.  This problem is especially
important for adaptive-optics (AO) imagery.  Presently, all working AO
systems can only correct for atmospheric blurring at the location
of the guide star or laser beacon.  As one moves away from this location
in angle, the correction will degrade rapidly.\cite{stein}  Traditionally there
are two approaches for the deconvolution or analysis of such images.

\begin{enumerate}
\item The easiest approach is to sweep the problem under the rug!  While
it may be facetious to suggest that this as a real solution, it is true
that there may be some cases where the PSF structural variations
are significant in general, but may be ignored for specific problems.
One might imagine a diffraction-limited system in which the form
of the sharp core varied over the image, but the broad wings more or
less stayed the same; for some problems, the latter component of the
PSF may be more important.

\item Break the image up into many sub-domains, over which the PSF may
be regarded as constant.  This procedure is tedious at best, may
produce discontinuities when the sub-images are stitched back together,
and still must find a way to represent the PSF satisfactorily at
all the locations demanded.

\end{enumerate}

In this paper, I will outline a different approach that incorporates the
continuous variation of the PSF into the deconvolution, but that also uses
efficient full-image Fourier convolution; in this regard, it should be superior
to, but yet simpler than the sub-domain approach.
As it happens, this method simply strings together ideas already in the
literature for compactly modeling the PSF spatial variations, as well as
encoding the variations in a form that allows for efficient convolution.
Folding these methods into standard Lucy\cite{lucy}-Richardson\cite{rich}
deconvolution allows that algorithm to work quickly with a variable PSF;
this method is readily adopted, as it actually only incorporates
forward convolutions of the PSF (and its transpose) to estimate the
deconvolved image.

An important caveat is that while a spatially-variant PSF
deconvolution algorithm attempts to use the most accurate PSF possible
for any point in the image, this does not imply that 
the deconvolved image will have uniform resolution.  Although a forward
convolution of a source model to match an image with large PSF variations
may be readily done, with real image noise the resolution gains
offered by deconvolution of such images will always be tied to the
intrinsic resolution of the local PSF.
In the single-point AO case, for example, one will not be able to
deconvolve the entire image to the same fine resolution available
at the correction-point.  Variable-PSF deconvolution
cannot undo any real loss of structural information that occurs as the
resolution degrades away from the reference point.

\section{THEORY}

\subsection{Encoding the PSF Spatial Variability}

The critical step in deconvolution of a spatially-variant PSF is
to understand how ordinary convolution can be done efficiently
with a variable PSF.  This methodology can then just be folded in to
those deconvolution algorithms, such as Lucy-Richardson, that are based
on forward convolution.  This discussion is thus limited to how to
treat the variable PSF, rather than the actual deconvolution algorithm itself.

If we ignore noise for this discussion, then an astronomical image, $I,$
can be presented schematically as the convolution of an astronomical
source, $S,$ and a PSF, $P:$
\begin{equation}
I=S*P.
\end{equation}
In the general case of a PSF that varies with location over the image,
this equation can be expanded explicitly as:
\begin{equation}
I(x,y)=\int_{-\infty}^\infty\int_{-\infty}^\infty S(u,v)
~P(u,v,x-u,y-v)~du~dv,
\end{equation}
where $\{x,y\}$ are any pixel location in the image and $\{u,v\}$ are
the pixel locations in the source.  The PSF is assumed to have
a form that changes with $\{u,v\}.$
The key step is to recast the PSF as a sum of orthogonal functions,
each of which is constant over the domain, but whose varying amplitudes
serve to encode the PSF spatial variance.  This gives
\begin{equation}
\label{psf}
P(u,v,x,y)=\sum_{i=1}^\infty a_i(u,v)~p_i(x,y),
\end{equation}
where $\{p_i\}$ are of the orthogonal PSF components, and $\{a_i\}$ is a field
that specifies how the amplitudes of $\{p_i\}$ varies over the image
domain to describe overall variations in the PSF.
This representation allows the variables to be separated in the convolution,
\cite{stock,alard,kepner}
\begin{equation}
\label{conv}
I(x,y)=\sum_{i=1}^\infty \int_{-\infty}^\infty\int_{-\infty}^\infty
S(u,v)~a_i(u,v)~p_i(x-u,y-v)~du~dv.
\end{equation}
In other words, each PSF component can be convolved with the source
weighted by a coefficient field
over its full domain, which is easily done in the fourier domain, to make
an image component.  The image components are then summed
to make the final complete image.  Provided that the $\{a_i\}$
are known in advance, the computational burden of tracking the PSF
variability scales only linearly with the number of PSF components
over that required for a constant-PSF convolution.

\subsection{Karhunen-Lo\`eve PSF Decomposition}

Choice of the appropriate PSF basis functions, $\{p_i\},$ is critical
to the ease of computing $\{a_i\}$ as well minimizing as the computational
overhead associated with computing equation (\ref{psf}).  Of the large set
of basis functions that one might contemplate, Karhunen-Lo\`eve
decomposition of the PSF appears to offer an attractive solution
to both problems; it is presently used by the Sloan Digital Sky Survey
to characterize PSF variations\cite{lupton}.
Karhunen-Lo\`eve decomposition is closely related
to principal-component analysis.  It works by finding the most important
eigenvectors that describe the space spanned by a set of objects.
It is non-parametric and by maximizing the variance accounted for
by each eigenvector, effectively minimizes the total dimension of the basis.

Calculation of the Karhunen-Lo\`eve decomposition is simple.  If one
has a set of $N$ PSF observations, $P^*_i,$ distributed of an image
domain, one first normalizes the PSFs to a common integral, spatial
extent, and photocenter to remove such trivial variations from the
problem.  A covariance matrix between all the PSFs is then computed, with
\begin{equation}
C_{ij}=<P^*_i\ P^*_j>.
\end{equation}
The eigenvalues, $\lambda_i,$ of this matrix are then found;
the corresponding eigenvectors, $\{x_i\},$ define the PSF basis functions or
``eigen-PSF'' images:
\begin{equation}
p_i=\sum_{1=1}^Nx_{ij}P^*_j,
\end{equation}
where $x_{ij}$ is component $j$ of $x_i.$
One thus directly derives the basis function from the data itself
without recourse to any parametric forms.
The importance of any $x_i$ for characterizing the span of $\{P^*\}$
is given by the amplitude of its corresponding eigenvalue.
If the $\{p_i\}$ are sorted by the eigenvalue amplitude, then
the one can assess how many terms are really required to explain
the significant structure in the PSF observations.  Use of all $N$
eigen-PSFs will fit each observation exactly, including noise.
It is likely instead that a subset of $K$ eigen-PSFs will be sufficient.
The attractiveness of the Karhunen-Lo\`eve decomposition method
is that since it each eigen-PSF corresponds to a principal axis in
the space spanned by the PSF observations, some of the less significant
dimensions may be neglected, potentially allowing for $K<<N.$

With the Karhunen-Lo\`eve decomposition in hand, one immediately
has the decomposition of $\{P^*\},$ into the form specified by
equation (\ref{psf}), with the exception that the sum is completed at $N$
terms, rather than being infinite; as noted, it may be truncated further
by discarding nonsignificant eigen-PSFs.  The final step, is
to model the behavior of $\{a_i\}$ over the image domain so they
can expressed as continuous coefficient fields.  Unlike the
decomposition of the PSF observations, themselves, this step may
require fitting a parametric form or model to the coefficients
as a function of position within the image domain.  Choice of this
model may be driven by the context.  For AO observations, for example,
the PSFs broaden as function of radial distance from the reference
point and may have anisotropic shapes that are related to the
direction vector to the reference point.  In this case, it's natural
to express the coefficient fields in functions of polar coordinates.

The main caveat with the use of Karhunen-Lo\`eve decomposition coupled
to a model of the coefficient fields is that it provides only an
approximation to the true PSF at any point in the image domain.  Since
in general, one is likely to use some observational representation
of the PSF for any deconvolution problem, then it is possible that
this issue in this context presents no more additional burden that
that always required to understand how dependent the analysis of
the deconvolved image is on errors in the PSF.  That said, there are
two concerns specifically related to use of Karhunen-Lo\`eve decomposition.
\begin{enumerate}
\item For many deconvolution algorithms, the PSF must be positive
definite.  Even if a truncated Karhunen-Lo\`eve decomposition is always
positive definite at the locations of the $\{P^*\},$ does not mean that
the coefficient fields guarantee this to be true at all other image
locations, particularly near the image margins, where the modeled
PSF is likely to be an extrapolation.  Solution to this problem,
if it occurs, may be related to the way the coefficients fields are
modeled.  In practice, if the span of $\{P^*\}$ is limited, then
the first eigen-PSF will strongly resemble an average PSF, with
the next eigen-PSFs characterizing variations about this form sufficiently
limited that the total PSF remains positive definite.

\item A more important problem is that the Karhunen-Lo\`eve decomposition
is certainly not likely to preserve the unit-integral of the PSFs.
Fortunately this problem is simply solved by normalizing equation (\ref{psf})
by the convolution of the PSF with a field of unit amplitude.\cite{alard}
This gives the 
Karhunen-Lo\`eve approximated PSF at any location as
\begin{equation}
P_{KL}(u,v,x,y)=\sum_{i=1}^K a_i(u,v)~p_i(x,y)~\bigg/~\sum_{i=1}^K~
\int_{-\infty}^\infty \int_{-\infty}^\infty~a_i(u,v)~p_i(x-u,y-v)~du~dv.
\end{equation}
In practice, the denominator will be a single ``normalization image'' that
can either be used to scale the coefficient fields in advance, or
applied to any convolution after the fact as a sort of flat-field.
\end{enumerate}

\subsection{The Lucy-Richardson Deconvolution Algorithm}

Lucy-Richardson deconvolution works by iteratively comparing an estimate
of the intrinsic astronomical source with its image.
The first iterative step begins can begin with any initial estimate
of the source structure based on prior information, although in
practice most users of this method start with a constant level
over the entire source domain.  At any iterative cycle, $n,$ the
running estimate of the source is convolved with the PSF
\begin{equation}
I_n=S_n*P,
\end{equation}
to make an estimate of the true image.  A correction image is then
estimated by convolving a ratio of the true to estimated image by
the transpose of the PSF,
\begin{equation}
C_n=\left({I\over I_n}\right)*P^T.
\end{equation}
This is used to update the source estimate by multiplication,
\begin{equation}
S_{n+1}=C_nS_n
\end{equation}
For each Lucy-Richardson iteration thus requires one convolution of the
PSF with the source, and one convolution of the image to estimated-image
ratio with the PSF transpose.  The first step has already been given
in equation (\ref{conv}); the second step only requires substituting $\{p_i^T\}$
for $\{p_i\},$ noting that this step still uses the same $\{a_i\}.$
Using Lucy-Richardson for a variable PSF is thus conceptually simple,
although it does require more complex ``book keeping'' and indexing
to keep track of the various sets of PSF components, coefficient fields,
and their normalization.
%-------------
\begin{figure}
\begin{center}
\begin{tabular}{c}
\includegraphics[height=9cm]{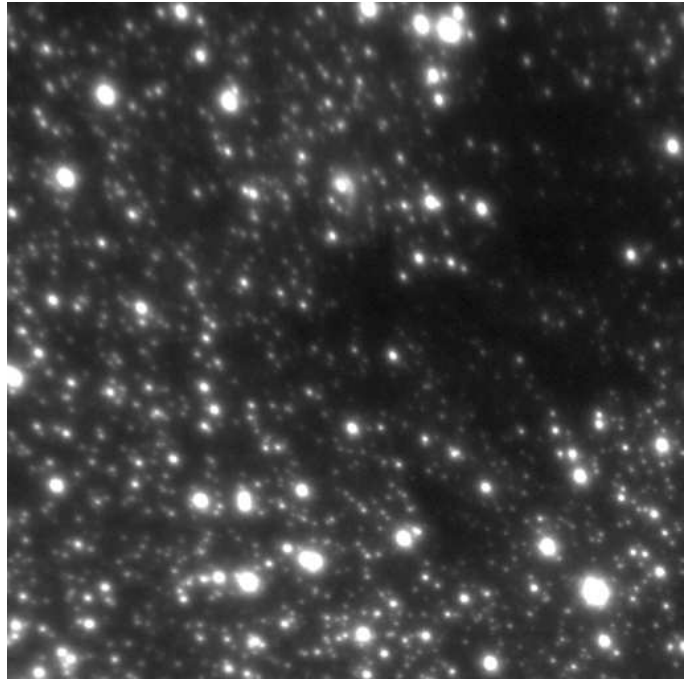}
\end{tabular}
\end{center}
\caption[example]
{\label{fig:gc2} An H-band image of a galactic center starfield
obtained with the QUIRC/Hokupa'a AO camera is shown.  It was obtained
under the Gemini Galactic Center Demonstration Science Program
and is publicly available from the Gemini website {\tt http://www.gemini.edu}.
The AO guidestar
is just outside the right edge of the image near the lower corner.
The area shown is $20.7\times20.7$ arcsecond.  A linear
stretch has been used.}
\end{figure}
%-------------

\section{DEMONSTRATION}

\begin{figure}
\begin{center}
\begin{tabular}{c}
\includegraphics[height=3cm]{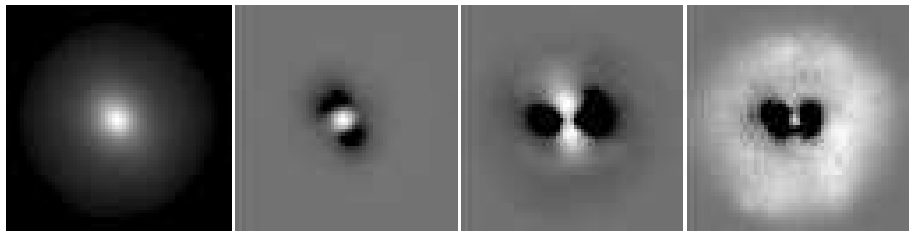}
\end{tabular}
\end{center}
\caption[example]
{\label{fig:eigenpsf} The first four eigen-PSFs for the galactic center
starfield are shown.  The most significant eigen-PSF is at the left,
and is positive definite.  The next eigen-PSFs are shown in order
of decreasing significance.  They have both positive and negative
components; the zero level is shown at the margins.  Stretches are
arbitrary for each eigen-PSFs and thus do not reflect their relative
significance.  Each subimage is 1.28 arcseconds on a side.}
\end{figure}

\begin{figure}
\begin{center}
\begin{tabular}{c}
\includegraphics[height=4cm]{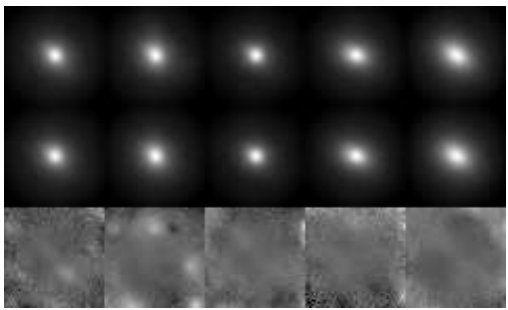}
\end{tabular}
\end{center}
\caption[example]
{\label{fig:mos} The top line shows images of five of the sample PSFs
and the middle line shows the model PSFs using just the first four
eigen-PSFs.  The bottom line of subimages shows the observed-model
differences.}
\end{figure}

Figure \ref{fig:gc2} shows an H-band images of a star field close to the
galactic center obtained with the QUIRC/Hokupa'a AO
Camera at the Gemini-N 8m telescope.  The AO guidestar is located
near the lower-right corner of the image.  Stars at this location
are compact and round, but become more elongated and broader
as the angular offset from the guidestar increases.
Several of the
brighter stars were drawn from the image to characterize
the PSF variation.  The Karhunen-Lo\`eve decomposition produces
the eigen-PSFs shown in Figure \ref{fig:eigenpsf}.  In general, the eigen-PSFs
%-------------
%-------------
need not correspond to any simple description of how the PSFs vary,
but in this case the role of the most important of the set are fairly
obvious.  The dominant eigen-PSF, as discussed above, looks very
much like a simple average over the stars.  The second eigen-PSF
is clearly moderating the width of the PSF, but because for this
image width correlates with elongation, it is not circularly
symmetric as a simple pure-width term would be.  In passing, this raises
an important caveat in the use of Karhunen-Lo\`eve decomposition.
Since the eigen-PSFs are determined directly by the span of the data,
they will vary as the PSF sample is varied.  Continuing on, the
third eigen-PSF appears to control the orientation of the PSF, while
the final eigen-PSF shown in part appears to be describing the PSF wings at
large angle.  The next eigen-PSFs (not shown) become more complex, but
less significant and begin describing the image noise of the PSF samples.
Figure \ref{fig:mos} shows that using just four eigen-PSFs provides
an excellent description of the PSFs and their variability.
%-------------
%-------------

%-------------
\begin{figure}
\begin{center}
\begin{tabular}{c}
\includegraphics[height=9cm]{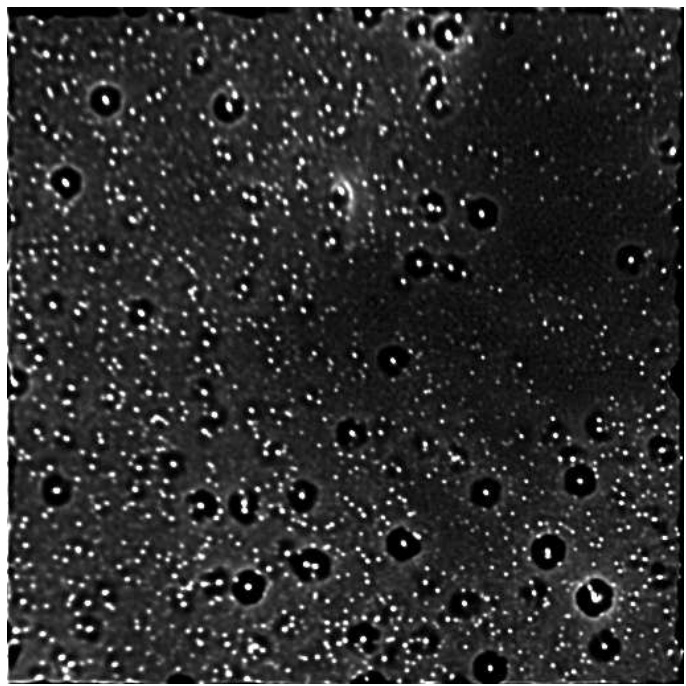}
\end{tabular}
\end{center}
\caption[example]
{\label{fig:gcd} The field shown in Figure \ref{fig:gc2} has been
deconvolved with 160 Lucy-Richardson iterations, using the
spatially-variant PSF.  The stretch is the same.}
\end{figure}
%-------------

The Lucy-Richardson deconvolution (160 iterations)
of the galactic center star-field
is shown in Figure \ref{fig:gcd},
starting with a constant image as the initial starting point.
One of the attractive feature of the Lucy-Richardson method is that
since it's based on simple arithmetic operations in the image
and Fourier domain, it's simple to implement at the high level
in any image processing system that emphasizes easy use of such
atomic operations.  I've long used Lucy-Richardson only
implemented as a high-level command-language program in the Vista
image processing system.  Augmentation of this program to implement
a spatially-variant PSF deconvolution of Lucy-Richardson was simple;
speed is emphasized over memory resources by preserving the
Fourier transforms of the eigen-PSFs and their transposes, which
only need to be calculated once during the initialization
of the deconvolution.

\begin{figure}
\begin{center}
\begin{tabular}{c}
\includegraphics[height=9cm]{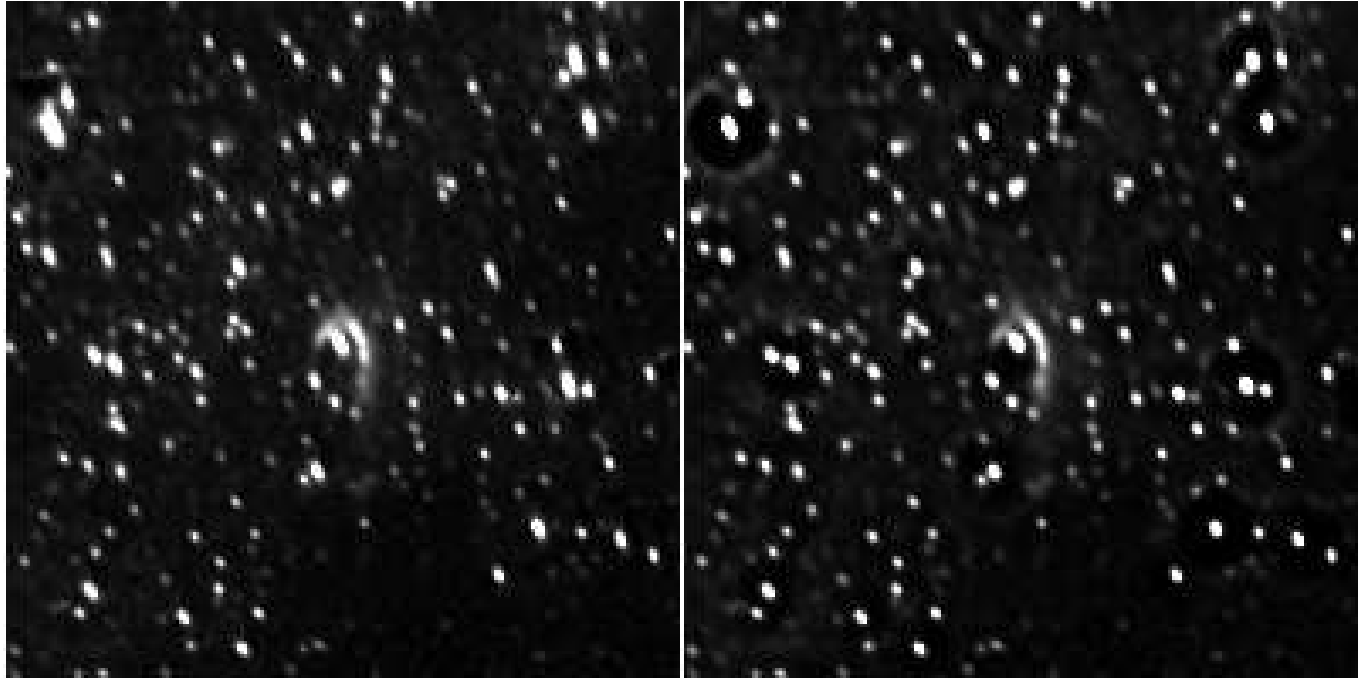}
\end{tabular}
\end{center}
\caption[example]
{\label{fig:comp}  This figure shows shows the deconvolution of the
portion of the field centered on the IRS8 source (a star with an
associated ISM bowshock) to demonstrate the differences between
two deconvolutions done on the assumption of a constant PSF (left)
and spatially-variant PSF (right).}
\end{figure}

At this writing, I'm just beginning to experiment with this method,
but Figure \ref{fig:comp} is offered as a comparison of the
difference between assuming a constant PSF, and allowing it to vary.
The deconvolution on the left used a PSF obtained close to the
AO guide star, the image on the right allows the PSF to vary.
The right image shows greater amplification of the stars, but
a reduction in their strong elongation.  Again, emphasizing
the point made in the introduction, this elongation cannot be
completely erased, but by using a PSF that is appropriate for
the field is has been reduced as well as the native Lucy-Richardson
algorithm will allow.  Future work will be to apply this
method to other AO datasets and {\it HST} images as well.

%-------------
%-------------
\acknowledgments

I thank Alex Szalay and Robert Lupton for introducing me to the use of the
Karhunen-Lo\`eve transform to represent PSFs.  I thank Dave De Young
for useful conversations.

\end{document}